\documentclass[english,aps,pra,reprint,noshowpacs,groupedaddress]{revtex4-1}  
\usepackage{graphicx,tabularx,times,hyperref}
\usepackage[T1]{fontenc}	
\usepackage[latin9]{inputenc}	
\usepackage{geometry}		
\usepackage{graphicx}
\usepackage[above,below]{placeins}	
\usepackage{times}
\usepackage{hyperref}  
\hypersetup{colorlinks=true,urlcolor=blue,citecolor=blue,linkcolor=blue}   
\setcitestyle{numbers,square}
\usepackage{array}
\usepackage{amsmath}
\usepackage{units}
\usepackage{graphicx}
\usepackage{braket}
\usepackage{xfrac}
\usepackage{enumitem}
\usepackage[normalem]{ulem}
\usepackage{booktabs}
\usepackage{tabularx}
\usepackage{tikz}
\usepackage{multirow, bigstrut}
\usepackage{hhline}
\usepackage{times}

\usetikzlibrary{shapes.geometric, arrows, matrix, positioning}
\usetikzlibrary{fit}
\tikzset{%
highlight/.style={rectangle,rounded corners,draw,
fill opacity=0.5,thick,inner sep=0pt}
}

\newcommand\Circle[1]{%
\tikz[baseline=(char.base)]\node[circle,draw,inner sep=2pt] (char) {#1};}

\usepackage{hyperref}

\hypersetup{
colorlinks=true,
linkcolor=cyan,
filecolor=magenta, 
urlcolor=blue,
}

\begin{document}


\title{Student difficulties with finding the corrections to the energy spectrum of the hydrogen atom for the strong and weak field Zeeman effects using degenerate perturbation theory}
\author{Emily Marshman}
\author{Christof Keebaugh}
\author{Chandralekha Singh}
\affiliation{Department of Physics and Astronomy, University of Pittsburgh, Pittsburgh, PA 15260}

\begin{abstract}

We discuss an investigation of student difficulties with the corrections to the energy spectrum of the hydrogen atom for the strong and weak field Zeeman effects using degenerate perturbation theory. This investigation was carried out in advanced quantum mechanics courses by administering written free-response and multiple-choice questions and conducting individual interviews with students. We discuss the common student difficulties related to these concepts which can be used as a guide for creating learning tools to help students develop a functional understanding of concepts involving the corrections to the energy spectrum due to the Zeeman effect. 
\end{abstract}

\maketitle

\section{INTRODUCTION AND BACKGROUND}

The Zeeman effect in the hydrogen atom is the shift in the energy spectrum due to the presence of a magnetic field, 
and it is proportional to the strength of the magnetic field. 
Here, we focus on two limiting cases: the strong and weak field Zeeman effects. The strong field Zeeman effect occurs when the corrections to the energies due to the Zeeman term are much greater than the corrections to the energies due to the fine structure term. The weak field Zeeman effect occurs when the corrections to the energies due to the fine structure term are much greater than the corrections to the energies due to the Zeeman term.
The Time-Independent Schr\"{o}dinger Equation (TISE) for the Hamiltonian with the fine structure and Zeeman corrections cannot be solved exactly. Nevertheless, 
since the fine-structure term and, in general, the Zeeman term are significantly smaller than the unperturbed Hamiltonian, perturbation theory (PT) is an excellent method for determining the approximate solutions to the TISE and corrections to the energy spectrum of the hydrogen atom. 
Due to the degeneracy in the hydrogen atom energy spectrum, degenerate perturbation theory (DPT) must be used to find the corrections for the strong and weak field Zeeman effect. 

It is important to help students develop a functional understanding of DPT in order to find the corrections to the energies for the strong and weak field Zeeman effects. However, quantum mechanics (QM) is particularly challenging for upper-level undergraduate and graduate students [1-9]
and students often struggle with DPT. Therefore, we investigated student difficulties with finding the first-order corrections to the energies of the hydrogen atom for the strong and weak field Zeeman effects using DPT. 

We first discuss the requisite knowledge students must have to use DPT in general and in the contexts of the strong and weak field Zeeman effects. 
PT is a useful approximation method for finding the energies and the energy eigenstates for a system for which the TISE is not exactly solvable. The Hamiltonian $\hat{H}$ for the system can be expressed as the sum of two terms, the unperturbed Hamiltonian $\hat{H}^0$ and the perturbation $\hat{H}'$, i.e., $\hat{H}=\hat{H}^0+\hat{H}'$. The TISE for the unperturbed Hamiltonian, $\hat{H}^0\psi_n^0 = E_n^0\psi_n^0$, is exactly solvable where $\psi_n^0$ is the $n^{th}$ unperturbed energy eigenstate and $E_n^0$ is the unperturbed energy. 
The energies can be approximated as $E_n =E_n^0 + E_n^1+E_n^2 + \ldots$ where $E_n^i$ for $i=1,2,3..$ are the $i^{\textrm th}$ order corrections to the $n^{\textrm th}$ energy of the system. 
In PT, the first-order corrections to the energies are
$E_n^1 = \langle \psi_n^0|\hat{H}'|\psi_n^0\rangle$
and the first-order corrections to the unperturbed energy eigenstates are 
$|\psi_n^1\rangle = \sum_{m \neq n} \frac{\langle \psi_m^0|\hat{H}'|\psi_n^0\rangle }{(E_n^0-E_m^0)}|\psi_m^0\rangle$, in which
$\left\{|\psi_n^0\rangle \right\}$ is a complete set of eigenstates of the unperturbed Hamiltonian 
$\hat{H}^0$. 
If the eigenvalue spectrum of $\hat{H}^0$ has degeneracy, 
the corrections to the energies and energy eigenstates are only valid provided one uses a {\it good} basis. 
For a given $\hat{H}^0$ and $\hat{H}'$, a {\it good} basis consists of a complete set of eigenstates of $\hat{H}^0$ that diagonalizes $\hat{H}'$ in each degenerate subspace of $\hat{H}^0$.

For a hydrogen atom in an external magnetic field, one can use DPT 
to find the corrections to the energy spectrum. 
Using standard notations, the unperturbed Hamiltonian $\hat{H}^0$ of a hydrogen atom 
is $\hat{H}^0 = \frac{\hat{p}^2}{2m}-\frac{e^2}{4 \pi \epsilon_0}\left(\frac{1}{r} \right)$, which accounts only for the interaction of the electron with the nucleus via Coulomb attraction. The solution for the TISE for the hydrogen atom with Coulomb potential energy gives the unperturbed energies $E_n^0=-\frac{13.6\textrm{eV}}{n^2}$, where $n$ is the principle quantum number. The perturbation is $\hat{H}' =\hat{H}'_{fs}+\hat{H}'_Z$, in which $\hat{H}'_{Z}$ is the Zeeman term and $\hat{H}'_{fs}$ is the fine structure term. 
We note that for each $n$ (i.e., each degenerate subspace of  $\hat{H}^0$), $\hat{H}^0$ for the hydrogen atom is 
diagonal when any complete set of orthogonal states is chosen for the angular part of the basis. Thus, so long as the radial part of the wavefunctions  
corresponding to the eigenstates of $\hat{H}^0$ is chosen as the basis, the choice of a good basis amounts to choosing the angular part of the basis (the part of the basis that reflects both orbital and spin angular momentum) appropriately. Therefore, for each $n$, we focus on the angular part of the basis to find a good basis for the perturbation $\hat{H}'$ corresponding to the fine structure and Zeeman corrections to the hydrogen atom.
For the angular part of the basis for each $n$, states in the coupled representation $|l,j, \ m_j \rangle$ are labeled by the quantum numbers  $ l, \ s, \ j$, and $m_j$ and the total angular momentum is defined as $\vec{J}=\vec{L}+\vec{S}$ (all notations are standard and $s=1/2$ has been suppressed from the states $|l,j, \ m_j \rangle$ since $s=1/2$ is a fixed value for a hydrogen atom).  
On the other hand, states {\small $|l,\ m_l,\ m_s \rangle$} in the uncoupled representation are labeled by the quantum numbers  $ l, \ s, \ m_l,$ and $m_s$  (notations are standard).

In the limiting cases of the strong and weak field Zeeman effect, the perturbation $\hat{H}'$ can be separated into two terms $\hat{H}'=\hat{H}'_{strong}+\hat{H}'_{weak}$, in which $\hat{H}'_{strong}$ is the stronger perturbation and $\hat{H}'_{weak}$ is the weaker perturbation. The corrections to the energies due to the stronger perturbation $\hat{H}'_{strong}$ are larger than the corrections due to the weaker perturbation $\hat{H}'_{weak}$. In these limiting cases, in order to find the corrections to the energies,
one useful approach is to use DPT via a two-step approximation. In the first step, the stronger perturbation $\hat{H}'_{strong}$ is treated as the only perturbation. A good basis for step 1 is one that diagonalizes the unperturbed Hamiltonian $\hat{H}^0$ and also diagonalizes the stronger perturbation $\hat{H}'_{strong}$ in each degenerate subspace of the unperturbed Hamiltonian $\hat{H}^0$. After a good basis has been identified for step 1, the first order corrections for the stronger perturbation $\hat{H}'_{strong}$ are determined. In the second step of the two-step approximation, $\hat{H}^0_{strong}=\hat{H}^0+\hat{H}'_{strong}$ is the new unperturbed Hamiltonian and the weaker perturbation $\hat{H}'_{weak}$ is treated as the perturbation. For step 2, a good basis is one that diagonalizes the unperturbed Hamiltonian $\hat{H}^0_{strong}$ and also diagonalizes $\hat{H}'_{weak}$ in each degenerate subspace of $\hat{H}^0_{strong}$. Once a good basis for step 2 has been identified, the first order corrections to the energies due to the weaker perturbation can be determined. The total first-order corrections to the energies are the sum of the corrections from steps 1 and 2.


The following steps describe how to determine a good basis and the first order corrections to the energies for the strong field Zeeman effect: (1) Treat the stronger perturbation $\hat{H}'_Z$ as the only perturbation on the unperturbed Hamiltonian $\hat{H}^0$, identify that a basis consisting of states in the uncoupled representation forms a good basis for the unperturbed Hamiltonian $\hat{H}^0$ and the stronger perturbation $\hat{H}'_{Z}$ (since $\hat{H}^0$ is diagonal in the uncoupled representation and $\hat{H}'_Z$ is diagonal in each degenerate subspace of $\hat{H}^0$ in the uncoupled representation), and determine the first-order corrections to the energies due to the stronger perturbation $\hat{H}'_Z$; (2) Treat the weaker perturbation $\hat{H}'_{fs}$ as the perturbation on $\hat{H}^0_{Z}=\hat{H}^0+\hat{H}'_{Z}$, identify that a basis consisting of states in the uncoupled representation forms a good basis for the unperturbed Hamiltonian $\hat{H}^0_Z$ and the weaker perturbation $\hat{H}'_{fs}$ (since $\hat{H}^0_Z$ is diagonal in the uncoupled representation and $\hat{H}'_{fs}$ is diagonal in the degenerate subspaces of $\hat{H}^0_Z$ in the uncoupled representation), and determine the first-order corrections to the energies due to the weaker perturbation $\hat{H}'_{fs}$; (3) The sum of the first-order corrections obtained in steps 1 and 2 are the first-order corrections to the energy spectrum of the hydrogen atom.
For the weak field Zeeman effect, the dominant fine structure term is the only perturbation on $\hat{H}^0$ in step 1 and the weaker perturbation $\hat{H}'_{Z}$ is the perturbation on the Hamiltonian $\hat{H}^0_{fs}=\hat{H}^0+\hat{H}'_{fs}$ in step 2.  In the weak field Zeeman effect, the coupled representation forms a good basis for both step 1 and 2.

\section{METHODOLOGY}
Student difficulties with the corrections to the energies of the hydrogen atom for the strong and weak field Zeeman effects using DPT were investigated using two years of data involving responses from 52 upper-level undergraduate students and 42 first-year graduate students to open-ended and multiple-choice questions administered after traditional instruction in relevant concepts. The undergraduates were in an upper-level undergraduate QM course, and graduate students were in a graduate-level QM course. Additional insight about the difficulties was gained from 13 individual think-aloud interviews (a total of 45 hours).  
Students were provided with all relevant information discussed in the introduction and background section and had lecture-based instruction in relevant concepts.  Similar percentages of undergraduate and graduate students displayed difficulties with DPT. 

After analyzing responses of 32 undergraduates on similar questions administered in two previous years, we posed the following question to 20 undergraduate and 42 graduate students in the following two years as part of an in-class quiz after traditional lecture-based instruction to examine student difficulties  
(in which the strong field and weak field Zeeman effects were listed individually in two separate questions):\\
\noindent{\it {\bf Q1.} A perturbation $\hat{H}'=\hat{H}'_{fs}+\hat{H}'_{Z}$ acts on a hydrogen atom with the unperturbed Hamiltonian $\hat{H}^0 = -\frac{\hbar^2}{2m} \nabla^2 - \frac{e^2}{4 \pi \epsilon_0}\left(\frac{1}{r}\right)$. For the perturbation $\hat{H}'=\hat{H}'_{fs}+\hat{H}'_{Z}$, circle \underline{\bf ALL} of the representations that form a good basis for the strong and weak field Zeeman effect and explain your reasoning. Assume that for all cases the principal quantum number $n=2$.
 \\
i. Coupled representation,\\
ii. Uncoupled representation,\\
iii. {\bf ANY} arbitrary orthonormal basis constructed with a linear combination of states in the coupled representation,\\
iv. {\bf ANY} arbitrary orthonormal basis constructed with a linear combination of states in the uncoupled representation,\\
v. Neither coupled nor uncoupled representation.}

The correct answer for the strong field Zeeman effect is option ii 
and the correct answer for the weak field Zeeman effect 
is option i. 
Below, we discuss difficulties with corrections to the energies 
due to the strong and weak field Zeeman effects.  

\section{STUDENT DIFFICULTIES}

Students had several difficulties with DPT in general (not restricted to the context of the strong and weak field Zeeman effects only). For example, when students were asked to determine 
a good basis for finding the corrections to the energies of the hydrogen atom due to fine structure,  many students did not even realize that DPT should be used. Other students knew that they had to use DPT to find corrections to the wavefunction, but they did not use DPT to find the first-order corrections to the energies, incorrectly claiming that DPT was not needed since no terms in $E_n^1 = \langle \psi_n^0|\hat{H}'|\psi_n^0\rangle$) ``blow up''. 
  
Moreover, even if students realized that DPT should be used for the strong and weak field Zeeman effects, many of them admitted that they had memorized which representation was a good basis in a given situation.  
Memorization of which basis to use often masked the fact that students did not have a deep understanding of DPT.  
Table \ref{goodrep} shows that many students struggled to identify a good basis for finding corrections to the energy spectrum due to the strong and weak field Zeeman effects. Below, we discuss some specific student difficulties: 
\begin{table}[t]
\caption[Difficulties]{Percentages of undergraduate (U) ($N=20$) and graduate students (G) ($N=42$) who answered Q1 correctly.
 }
\label{goodrep}
\centering
\begin{tabular}
{|c|c|c|}
\hline
Limiting Case & U & G\\
\hline
Strong Field&40\%&29\%\\ 
\hline
Weak Field&25\%&31\%\\
\hline
\end{tabular} %
\end{table}

{\bf Not focusing on both $\hat{H}^0$ and $\hat{H}'$ when determining a good basis:}  Students with this type of difficulty focused on the bases that make $\hat{H}^0$ diagonal but did not give consideration to $\hat{H}'$ when finding a good basis. 
For example, in the first step of the two-step approximation for the weak field Zeeman effect, some students incorrectly claimed that the uncoupled representation forms a good basis because it diagonalizes the operator $\hat{H}^0$. 
Interviews suggest that these students often did not realize that $\hat{H}'_{fs}$  is not diagonal in each degenerate subspace of $\hat{H}^0$ if the uncoupled representation is chosen as a basis and the corrections using this representation will yield incorrect values inconsistent with experiments.

{\bf Focusing on the degeneracy in $\hat{H}'_{weak}$ instead of the degeneracy in $\hat{H}^0_{strong}$ when determining a good basis:}
When determining whether DPT should be used and whether a basis is a good basis, some students incorrectly focused on the degenerate subspaces of $\hat{H}'$ instead of $\hat{H}^0$. For example, when students were asked to find the energy corrections in the first step of the two-step approximation, some students incorrectly focused  on the degeneracy in $\hat{H}'_{strong}$ to determine whether DPT should be used and whether the basis provided was good. In particular, they focused on whether the degenerate subspaces in $\hat{H}'_{strong}$ were diagonal to determine if the basis was good (instead of whether $\hat{H}'_{strong}$ was diagonal in the degenerate subspaces of $\hat{H}^0$). 
An analogous student difficulty was also prevalent in step 2 of the two-step approximation. In particular, in order to determine whether a basis is a good basis for the strong or weak field Zeeman effect in step 2, students must identify the degenerate subspaces of $\hat{H}^0_{strong}=\hat{H}^0+\hat{H}'_{strong}$ and determine whether or not the weaker perturbation $\hat{H}'_{weak}$ is diagonal  in each degenerate subspace of $\hat{H}^0_{strong}$. However, many students incorrectly focused on the degeneracy and degenerate subspaces of $\hat{H}'_{weak}$ instead of the degenerate subspace of $\hat{H}^0_{strong}$ to determine if DPT should be used and if the basis provided was good. 

For example, during the portion of the interview regarding the strong field Zeeman effect, in step 2, students were given the strong field Zeeman  Hamiltonian $\hat{H}^0_Z=\hat{H}^0+\hat{H}'_Z$ from step 1 and the weaker perturbation $\hat{H}'_{fs}$ in matrix form in the uncoupled representation for $n=2$ (since the uncoupled representation is a good basis for step 1 of the two-step approximation method). The students were then asked to identify the $\hat{H}'_{fs}$ matrix in each degenerate subspace of $\hat{H}^0_Z=\hat{H}^0+\hat{H}'_Z$ and explain whether or not the uncoupled representation forms a good basis in step 2 of the 2-step approximation method.
In the $n=2$ subspace with $s=\frac{1}{2}$, the $\hat{H}^0_Z= \hat{H}^0 + \hat{H}'_{Z}$ matrix provided to students to probe their understanding is the following in which the basis states are chosen in the uncoupled representation ({\scriptsize $|l,\ m_l,\ m_s \rangle$}) in the order {\scriptsize $|0,\ 0, \ \frac{1}{2}\rangle$, $|0,\ 0, \ -\frac{1}{2}\rangle$,$|1,\ 1, \ \frac{1}{2}\rangle$, $|1,\ 1, \ -\frac{1}{2}\rangle $, $|1,\ 0, \ \frac{1}{2}\rangle$, $|1,\ 0, \ -\frac{1}{2}\rangle$, $|1,\ -1, \ \frac{1}{2}\rangle$}, and {\scriptsize $|1,\ -1, \ -\frac{1}{2}\rangle$} in which $\beta=\mu_B B_{ext}$ ($B_{ext}$ is a uniform, time-independent external magnetic field along the $\hat{z}$-direction and $\mu_B$ is the Bohr magneton): 
\resizebox{\columnwidth}{!}{

$
\left[ \begin{array}{cccccccc}
\boxed{E_2+ \beta}&0&0&0&0&0&0&0\\
0&\underline{E_2- \beta}&0&0&0&0&0&0\\
0&0&E_2+2 \beta&0&0&0&0&0\\
0&0&0&\Circle{$E_2$}&0&0&0&0\\
0&0&0&0&\boxed{E_2+ \beta}&0&0&0\\
0&0&0&0&0&\underline{E_2- \beta}&0&0\\
0&0&0&0&0&0&\Circle{$E_2$}&0\\
0&0&0&0&0&0&0&E_2-2 \beta\\
\end{array} \right]
$.}

{\small $\hat{H}^0_Z$} has three separate two-fold degeneracies for the energies {\small $E_2+\beta, E_2-\beta$}, and {\small $E_2$ }as indicated by the boxed, underlined, and circled matrix elements of $\hat{H}^0_Z$ above.  In order to determine whether a basis consisting of states in the uncoupled representation forms a good basis, $\hat{H}'_{fs}$ must be diagonal in each of these three degenerate subspaces of $\hat{H}^0_Z$.  The $\hat{H}'_{fs}$ matrix in the $n=2$ subspace in which the basis states are chosen in the same order as they were for the $\hat{H}^0_Z$ matrix above is shown below:
\begin{small}

$
\frac{(-13.6\textrm{ eV}) \alpha^2}{192}
\left[
\begin{array}{cccccccc}
\boxed{15}&0&0&0&\boxed{0}&0&0&0\\
0&15&0&0&0&0&0&0\\
0&0&3&0&0&0&0&0\\
0&0&0&11&4\sqrt{2}&0&0&0\\
\boxed{0}&0&0&4\sqrt{2}&\boxed{7}&0&0&0\\
0&0&0&0&0&7&4\sqrt{2}&0\\
0&0&0&0&0&4\sqrt{2}&11&0\\
0&0&0&0&0&0&0&3\\
\end{array}
\right]
$
\end{small}.

From the boxed matrix elements, $\hat{H}'_{fs}$ in the degenerate subspace of $\hat{H}^0_Z$ for the degenerate energy $E_2+\beta$ is 
\begin{tiny}
$\frac{(-13.6\textrm{ eV}) \alpha^2}{192}
\left[
\begin{array}{cc} 
15&0\\
0&7\\
\end{array}
\right]$
\end{tiny}.  
Similarly, one can determine $\hat{H}'_{fs}$ in the degenerate subspace of $\hat{H}^0_Z$ for the degenerate energies $E_2-2\beta$ and $E_2$ as the underlined and circled matrix elements, respectively.
However, students often did not realize that they should focus on the degeneracy of the Hamiltonian $\hat{H}^0_Z=\hat{H}^0+\hat{H}'_Z$ and instead they focused on the degeneracy of the weak perturbation $\hat{H}'_{fs}$ by examining the diagonal matrix elements of $\hat{H}'_{fs}$ that were equal. For example, they focused on the degenerate subspace \begin{tiny}
$\frac{(-13.6\textrm{ eV}) \alpha^2}{192}\left[\begin{array}{cc} 
15&0\\
0&15\
\end{array}
\right]$
\end{tiny} in $\hat{H}'_{fs}$.
In particular, they incorrectly focused on whether the degenerate subspaces of $\hat{H}'_{fs}$ were diagonal to determine whether a given basis is a good basis. However, the degeneracy of the weaker perturbation $\hat{H}'_{fs}$ is not relevant to determining a good basis. Instead, they should have identified the degenerate subspaces of $\hat{H}^0_Z=\hat{H}^0+\hat{H}'_Z$ and determined if the weaker perturbation $\hat{H}'_{fs}$ is diagonal in the degenerate subspaces of $\hat{H}^0_Z=\hat{H}^0+\hat{H}'_Z$ to conclude if a given basis is a good basis in step 2
 of the two-step process. 

{\bf Incorrectly claiming that $\hat{H}'_{weak}$ must be diagonal in each degenerate subspace of $\hat{H}^0$ in a good basis when using the two-step approximation:} 
Many students claimed that, in a good basis for step 2 of the two-step approximation,  $\hat{H}'_{weak}$ must be diagonal in the  degenerate subspace of $\hat{H}^0$ as opposed to the degenerate subspaces of $\hat{H}^0_{strong}$. They did not realize that when using the two-step approximation in the limiting cases, the weaker perturbation $\hat{H}'_{weak}$ need only be diagonal in each degenerate subspace of the stronger Hamiltonian $\hat{H}^0_{strong}$ determined in step 1 (as opposed to each degenerate subspace of $\hat{H}^0$). In the strong field Zeeman effect, a basis consisting of states in the uncoupled representation forms a good basis.
 Despite the fact that the weaker perturbation $\hat{H}'_{fs}$ is not diagonal in the degenerate subspace of $\hat{H}^0$, when the uncoupled representation is chosen as the basis, $\hat{H}'_{fs}$ {\it is} diagonal in each degenerate subspace of $\hat{H}^0_Z=\hat{H}^0+\hat{H}'_Z$ after accounting for the splitting of the energy levels due to the stronger pertubation $\hat{H}'_Z$. 
Many students struggled with the fact that the weaker perturbation $\hat{H}'_{fs}$ must only be diagonal in each degenerate subspace of $\hat{H}^0_Z$ in step 2.
For example, one interviewed student claimed ``the uncoupled is not a good basis (for strong field Zeeman effect) since $\hat{H}'_{fs}$ is not diagonal in the uncoupled representation.  So we will have off-diagonal (matrix) elements (of $\hat{H}'_{fs}$).''  

{\bf Not realizing that some of the degeneracy is broken after taking into account the stronger perturbation, 
allowing $\hat{H}'_{weak}$ to be diagonal in each degenerate 
subspace 
of $\hat{H}^0_{strong}$:}
Many students struggled with the fact that the utility of the two-step approximation for the strong and weak field Zeeman effects lies in the fact that some of the degeneracy is broken in step 1 of the two-step approximation when the stronger perturbation $\hat{H}'_{strong}$ is treated as the only perturbation on the unperturbed Hamiltonian $\hat{H}^0$.  They did not realize that in general, after taking into account the stronger perturbation in step 1, the dimension of some of the degenerate subspaces is reduced.  Therefore, in step 2 when $\hat{H}^0_{strong}=\hat{H}^0+\hat{H}'_{strong}$ is treated as the new unperturbed Hamiltonian, the degeneracy of $\hat{H}^0_{strong}=\hat{H}^0+\hat{H}'_{strong}$ is less than the degeneracy of $\hat{H}^0$, making it possible for the weaker perturbation $\hat{H}'_{weak}$ to be diagonal in the degenerate subspaces of $\hat{H}^0_{strong}=\hat{H}^0+\hat{H}'_{strong}$.
For example, in the strong field Zeeman effect, a basis consisting of states in the uncoupled representation forms a good basis for $\hat{H}^0$ and $\hat{H}_{Z}$ in step 1 and also in step 2. However, students often did not realize that for $n=2$, the degeneracy in the new unperturbed Hamiltonian $\hat{H}^0_Z=\hat{H}^0+\hat{H}'_Z$ is reduced to three separate two-fold degeneracies (instead of an 8-fold degeneracy in the unperturbed Hamiltonian $\hat{H}^0$). They did not realize that in step 2, in the uncoupled representation, the weaker perturbation $\hat{H}'_{fs}$ is diagonal in each of these $2\times 2$ subspaces of the Hamiltonian $\hat{H}^0_Z=\hat{H}^0+\hat{H}'_Z$ and the uncoupled representation is a good basis for finding corrections.

In interviews, students often argued that {\it neither} a basis consisting of states in the coupled representation nor a basis consisting of states in the uncoupled representation form a good basis even in the limiting cases since neither is a good basis for both the Zeeman term $\hat{H}'_{Z}$ and the fine structure term $\hat{H}'_{fs}$. They claimed that even in the limiting cases, one must find a basis that diagonalizes {\it both} the Zeeman term $\hat{H}'_{Z}$ and the fine structure term $\hat{H}'_{fs}$. Further probing suggests that they often did not realize that in the limiting cases some of the degeneracy is lifted after step 1 in the two-step process so that  the basis chosen in step 1 remains a good basis in step 2. 
\section{CONCLUSION AND FUTURE PLAN}
Both undergraduate and graduate students struggled with finding corrections to the energy spectrum of the hydrogen atom for the strong and weak field Zeeman effects using DPT. 
We used the difficulties as resources in developing a Quantum Interactive Learning Tutorial (QuILT) to help students develop a good grasp of these concepts.
The initial results from the QuILT are encouraging. 
\begin{acknowledgments}
We thank the NSF for award PHY-1505460. 
\end{acknowledgments}

\bibliographystyle{apsrev}

\end{document}